\documentstyle[12pt]{article}

\newcommand{\be}{\begin{equation}}
\newcommand{\ee}{\end{equation}}
\newcommand{\dlt}{\delta}
\newcommand{\Dlt}{\Delta}
\newcommand{\ra}{\rightarrow}
\newcommand{\vp}{\varphi}
\newcommand{\bt}{\beta}
\newcommand{\al}{\alpha}
\newcommand{\prt}{\partial}
\newcommand{\om}{\omega}
\newcommand{\lbd}{\lambda}
\newcommand{\ep}{\varepsilon}
\newcommand{\br}{{\bf r}}
\newcommand{\bk}{{\bf k}}

\begin{document}

\begin{center}
{\Large{\bf Kinetic energy of Bose systems and variation of statistical
averages} \\ [5mm]

V.I. Yukalov} \\ [5mm]

{\it
Bogolubov Laboratory of Theoretical Physics, \\
Joint Institute for Nuclear Research, Dubna 141980, Russia}

\end{center}

\vskip 2cm

\begin{abstract}

The problem of defining the average kinetic energy of statistical systems 
is addressed. The conditions of applicability for the formula, relating 
the average kinetic energy with the mass derivative of the internal energy, 
are analysed. It is shown that incorrectly using this formula, outside its 
region of validity, leads to paradoxes. An equation is found for a 
parametric derivative of the average for an arbitrary operator. A special 
attention is paid to the mass derivative of the internal energy, for which 
a general formula is derived, without invoking the adiabatic approximation 
and taking into account the mass dependence of the potential-energy operator. 
The results are illustrated by the case of a low-temperature dilute Bose gas.

\end{abstract}



\newpage

\section{Introduction}

There exists a known formula by Landau and Lifshitz [1] (see \S 15)
connecting the average kinetic energy of a statistical system with the mass
derivative of the internal energy. This formula has recently been used in
several papers for evaluating the average kinetic energy of low-temperature
dilute Bose gas. This gas is known to be well described by the Bogolubov
theory [2--5]. However, there is an ambiguity in calculating the average
kinetic energy in the dilute-gas approximation [6]. To avoid this ambiguity,
several authors have applied the Landau-Lifshitz formula for evaluating the
kinetic energy. Unfortunately, one usually forgets the conditions, under
which this formula was derived, because of which the formula employed outside
its region of applicability, leads to incorrect results.

To be concrete, let us recall this formula [1] together with the conditions
of its validity. One considers a statistical system described by a Hamiltonian
\be
\label{1}
\hat H =\hat K +\hat W \; ,
\ee
which is a sum of the kinetic-energy operator $\hat K$ and of the
potential-energy operator $\hat W$. The kinetic-energy operator is assumed
to have the standard form of an operator for nonrelativistic particles, when
$\hat K\propto 1/2m$, so that
\be
\label{2}
\frac{\prt\hat K}{\prt m} = -\; \frac{1}{m}\; \hat K \; ,
\ee
where $m$ is the particle mass. The potential-energy operator is supposed to
have no dependence on the mass, such that
\be
\label{3}
\frac{\prt\hat W}{\prt m} = 0 \; .
\ee
The total internal energy of the system is given by the statistical average
\be
\label{4}
E \equiv \; <\hat H>
\ee
of the Hamiltonian (1). Let the mass be varied under the fixed temperature 
$T$, volume $V$, and entropy $S$. Then one has the average kinetic energy
\be
\label{5}
<\hat K> \; = -m \left ( \frac{\prt E}{\prt m}\right )_S \; ,
\ee
which is the Landau-Lifshitz formula [1]. Here and in what follows all 
parametric derivatives are taken under fixed temperature and volume, because 
of which this is not specified each time to make the resulting expressions 
less cumbersome.

It is, however, important to stress that fixing the entropy implies the
{\it adiabatic approximation}. Really, the entropy, by its definition, is 
a function $S=S(T,m)$ of temperature and mass. Hence, fixing the entropy 
means that the temperature becomes a function $T=T(S,m)$ of the given entropy 
and mass. Therefore, the derivative $(\prt T/\prt m)_S\neq 0$ is nonzero.
That is, in general, it is impossible to simultaneously fix the temperature 
and entropy. The latter is admissible only for some extremal cases, say, 
when $T=0$ and, hence, $S=const$. This is why formula (5), in the general 
sense, corresponds to an adiabatic approximation, becoming exact solely
for those cases, when both $T$ and $S$ can be made constant.

But even when $T=0$, hence $S=const$, it is necessary to keep in mind 
other conditions, only under which formula (5) is valid. Going outside the
region of applicability of this formula can yield principally wrong results.
As an illustration, let us consider the uniform low-temperature dilute Bose 
gas described by the Bogolubov theory [2--5] (see also the recent review 
papers [7--9]). The particle interactions are characterized by a scattering
length $a_s$. We assume that there are no bound states, so that $a_s$ does 
not depend on the mass. At zero temperature and in the first order with 
respect to the interaction intensity, the internal energy per particle is 
the ground-state energy
$$
\frac{E}{N} = 2\pi\; \frac{\rho a_s}{m} \qquad (T=0) \; ,
$$
where $\rho\equiv N/V$ is the average particle density. It is a trivial 
task to check (see the original works [2--5], review papers [7--9], or
the textbook [10]) that the average potential energy for this case is
$<\hat W>=E$. At the same time, invoking formula (5), one gets 
$<\hat K>=E$. By definition (4), one has $E=<\hat K+\hat W>$. Since 
$<\hat W>=<\hat K>=E$, one comes to the equality $E=2E$, implying that 
$1=2$, which is, of course, senseless. Such a paradoxical conclusion 
stems from the fact that Eq. (3), which is a necessary condition for 
the validity of formula (5), has been forgotten. 

In the following Section 2, we derive a general equation for the 
derivative, with respect to any parameter, of a statistical average for 
an arbitrary operator. In this derivation, the adiabatic approximation 
is not involved. The obtained general formula is illustrated by 
considering the derivatives of the internal energy, including the case 
of the dilute Bose gas. In Section 3, we show how the average kinetic 
energy can be evaluated directly  from the Bogolubov theory [2--5].

\section{Parametric variation of statistical averages}

Consider a statistical system characterized by a Hamiltonian $\hat H$. 
In equilibrium, the average of an operator $\hat A$ is given by the 
statistical average
\be
\label{6}
<\hat A>\; \equiv {\rm Tr}\; \hat\rho\hat A \; ,
\ee
with the Gibbs statistical operator
$$
\hat\rho = \frac{e^{-\bt\hat H}}{{\rm Tr}e^{-\bt\hat H}} \qquad 
\left (\bt \equiv \frac{1}{T}\right ) \; .
$$
Throughout the paper, the Heisenberg representation for the operators
is kept in mind, and the system of units is used, where the Boltzmann 
and Planck constants are set to unity, $k_B=1$, $\hbar=1$.

Let the Hamiltonian depend on a parameter $\lbd$, such that $\hat H=
\hat H(\lbd)$, and the considered operator be also dependent on the 
same parameter, $\hat A=\hat A(\lbd)$. Taking the parametric derivative 
of two traced operators, say $\hat A$ and $\hat B$, one should keep in 
mind the trace property
$$
{\rm Tr}\; \hat A\hat B = {\rm Tr}\; \hat B\hat A = \frac{1}{2}\;
{\rm Tr}\left [\hat A,\; \hat B\right ]_+ \; ,
$$
where the notation for the anticommutator
$$
\left [\hat A,\; \hat B\right ]_+ \; \equiv \hat A\hat B +
\hat B\hat A
$$
is used. That is, the parametric derivative
$$
\frac{\prt}{\prt\lbd} \; {\rm Tr}\; \hat A\hat B = \frac{1}{2}\;
{\rm Tr} \left (\left [ \frac{\prt\hat A}{\prt\lbd}\; , \;
\hat B\right ]_+ + \left [ \hat A,\; \frac{\prt\hat B}{\prt\lbd}
\right ]_+ \right )
$$
has to always be represented in the symmetrized form.

For what follows, we shall also need the notation for the covariance of 
two operators,
$$
{\rm cov}\left (\hat A,\; \hat B\right ) \equiv \frac{1}{2}\; <\left [
\hat A,\; \hat B\right ]_+> - <\hat A><\hat B> \; .
$$
This is symmetric,
$$
{\rm cov}\left ( \hat A,\; \hat B\right ) = 
{\rm cov}\left ( \hat B,\; \hat A\right ) \; ,
$$
and possesses the property of additivity,
$$
{\rm cov}\left ( \hat A + \hat B,\; \hat C\right ) =
{\rm cov}\left ( \hat A,\; \hat C\right ) +
{\rm cov}\left ( \hat B,\; \hat C\right )\; .
$$
The diagonal covariance defines the operator dispersion
$$
\Dlt^2(\hat A) \equiv {\rm cov}\left ( \hat A,\; \hat A\right )
= \; <\hat A^2> - <\hat A>^2 \; .
$$

Taking the parametric derivative of an operator average, we shall 
always keep the temperature and volume fixed. But no restrictions 
on the entropy will be imposed. Then
\be
\label{7}
\frac{\prt}{\prt\lbd} <\hat A> \; = \; 
<\frac{\prt\hat A}{\prt\lbd}> - \; \bt\; {\rm cov}\left ( 
\hat A,\; \frac{\prt\hat H}{\prt\lbd} 
\right )\; .
\ee
Of particular importance is the case, when the operator $\hat A$ is
self-adjoint, representing an observable quantity, such that
$$
(\vp_k,\hat A\vp_p) =(\hat A\vp_k,\vp_p) \; .
$$
Let also the operator $\hat A$ commute with $\hat H$, thus representing
an integral of motion,
$$
\left [\hat A,\; \hat H\right ] = 0\; .
$$
Under these conditions, the operators $\hat H$ and $\hat A$ possess a
common set of eigenfunctions,
$$
\hat H\vp_k = E_k\vp_k \; , \qquad \hat A\vp_k = A_k\vp_k \; .
$$
Owing to the Hamiltonian parametric dependence $\hat H=\hat H(\lbd)$,
the eigenfunctions also depend on this parameter, $\vp_k=\vp_k(\lbd)$. 
For the normalized functions, one has
$$
\frac{\prt}{\prt\lbd}\; (\vp_k,\vp_k) = 0 \; , \qquad
(\vp_k,\vp_k)= 1 \; .
$$
It is straightforward to check that
\be
\label{8}
\frac{\prt}{\prt\lbd} \left (\vp_k,\hat A\vp_k\right ) =  
\left (\vp_k,\; \frac{\prt\hat A}{\prt\lbd}\;\vp_k\right ) \; .
\ee
Comparing Eqs. (7) and (8), and recalling that
$$
\lim_{T\ra 0} \; \frac{\prt}{\prt\lbd}<\hat A> \; = 
\frac{\prt}{\prt\lbd}\; (\vp_0,\hat A\vp_0) \; , \qquad
\lim_{T\ra 0} < \frac{\prt\hat A}{\prt\lbd}> \; = \left (\vp_0,\;
\frac{\prt\hat A}{\prt\lbd}\;\vp_0 \right ) \; ,
$$
with $\vp_0$ being the ground state, we see that for a self-adjoint integral 
of motion $\hat A$, one finds the limit
\be
\label{9}
\lim_{\bt\ra\infty} \; \bt\; {\rm cov} \left (\hat A,\;
\frac{\prt\hat H}{\prt\lbd}\right ) = 0 \; .
\ee
In particular, when $\hat A$ is $\hat H$, then for the ground-state energy
$$
E_0 \equiv \left ( \vp_0,\; \hat H\vp_0\right )
$$
one gets
$$
\frac{\prt E_0}{\prt\lbd} =\left ( \vp_0,\; 
\frac{\prt\hat H}{\prt\lbd}\vp_0\right ) \qquad (T=0)
$$ 
and the limit
$$
\lim_{\bt\ra\infty}\; \bt\; {\rm cov}\left (\hat H,\;
\frac{\prt\hat H}{\prt\lbd}\right ) = 0 \; .
$$

For the internal energy $E\equiv<\hat H>$ at any temperature $T\geq 0$, 
we obtain the general formula
\be
\label{10}
\frac{\prt E}{\prt\lbd} = \; < \frac{\prt\hat H}{\prt\lbd} > - \;
\bt\;{\rm cov} \left (\hat H, \; \frac{\prt\hat H}{\prt\lbd}\right ) \; .
\ee
Clearly, the second term in the right-hand side of Eq. (10) is not
necessarily zero at finite temperature $T>0$.

The first term in the right-hand side of Eq. (10) can be expressed through
the derivative
\be
\label{11}
\frac{\prt F}{\prt\lbd} = \; <\frac{\prt\hat H}{\prt\lbd} >
\ee
of the free energy
$$
F = - T\ln{\rm Tr}\; e^{-\bt\hat H} \; .
$$
Hence, Eq. (10) can be rewritten as
\be
\label{12}
\frac{\prt E}{\prt\lbd} = \frac{\prt F}{\prt\lbd} \; - \;
\bt\; {\rm cov} \left (\hat H,\; \frac{\prt\hat H}{\prt\lbd}\right ) \; .
\ee
Due to the equality $F=E-TS$, one has
$$
\frac{\prt F}{\prt\lbd} = \frac{\prt E}{\prt\lbd} \; - \; T\;
\frac{\prt S}{\prt\lbd} \; .
$$
Then another form of Eq. (10) is
\be
\label{13}
\frac{\prt E}{\prt\lbd} = \; < \frac{\prt\hat H}{\prt\lbd}> + \;
T\; \frac{\prt S}{\prt\lbd} \; .
\ee
From the comparison of Eqs. (12) and (13) it follows that
\be
\label{14}
\frac{\prt S}{\prt\lbd} = -\bt^2\; {\rm cov}\left (\hat H,\; 
\frac{\prt\hat H}{\prt\lbd} \right )\; .
\ee
The same expression (14) can be derived from the direct differentiation 
of the definition for the entropy $S\equiv-{\rm Tr}\hat\rho\ln\hat\rho$, 
for which we have
$$
\frac{\prt S}{\prt\lbd} = \bt\; {\rm cov}\left ( \ln\hat\rho, \;
\frac{\prt\hat H}{\prt\lbd} \right )\; .
$$
Substituting here
$$
\ln \hat\rho = \bt (F-\hat H ) \; ,
$$
we return to Eq. (14). According to the limit (9), one gets
$$
\lim_{T\ra 0} \; T\; \frac{\prt S}{\prt\lbd} = 0\; .
$$

Let us emphasize that the second terms in the right-hand sides of Eqs. 
(10), (12), and (13), generally, are not zero for finite temperatures. 
It is only in the adiabatic approximation, when one has
\be
\label{15}
\left ( \frac{\prt E}{\prt\lbd}\right )_S = \left (
\frac{\prt F}{\prt\lbd} \right )_S = \; < \frac{\prt\hat H}{\prt\lbd}>
\qquad \left ( \frac{\prt S}{\prt\lbd} = 0\right ) \; .
\ee
The situation here is the same as is mentioned in the Introduction. At 
finite temperatures, the entropy is a function $S=S(T,\lbd)$. Hence fixing 
the entropy requires that the temperature be a function $T=T(S,\lbd)$ of 
the parameter $\lbd$, so that $(\prt T/\prt\lbd)_S\neq 0$. In general, it 
is impossible to fix simultaneously both the temperature and entropy. This 
could be done either in some marginal cases or approximately.

Let us now take the particle mass $m$ as a parameter $\lbd$. Then, with
the derivative
\be
\label{16}
\frac{\prt\hat H}{\prt m} = -\; \frac{1}{m} \; \hat K +
\frac{\prt\hat W}{\prt m} \; ,
\ee
from Eq. (10) we find
\be
\label{17}
\frac{\prt E}{\prt m} = -\; \frac{1}{m}\; <\hat K> + 
<\frac{\prt\hat W}{\prt m} > + \; T \; \frac{\prt S}{\prt m} \; .
\ee
This formula is a generalization of the Landau-Lifshitz equation (5).
At zero temperature, this becomes
\be
\label{18}
\frac{\prt E}{\prt m} = -\; \frac{1}{m}\; <\hat K> + 
<\frac{\prt\hat W}{\prt m} > \qquad \left ( T=0\right )\; .
\ee
The second term in the right-hand side of Eq. (18), generally, is not
zero. It is only under the supposition (3), when Eq. (18) reduces to 
the Landau-Lifshitz formula (5).

An obvious case, when the condition (3) is not valid, is exactly the 
case of the dilute Bose gas, whose interaction energy is proportional 
to the interaction intensity
$$
\Phi_0 \equiv 4\pi\; \frac{a_s}{m} \; .
$$
The potential of pair interactions for the dilute gas is proportional 
to $\Phi_0$, whether the Fermi contact potential or the Huang effective 
potential are considered [10]. Therefore for the dilute gas, one has
\be
\label{19}
\frac{\prt\hat W}{\prt m} = -\; \frac{1}{m}\; \hat W \; .
\ee
Note that relation (19) is valid for many systems, but not solely for 
the dilute gas. Together with Eq. (2), this gives
\be
\label{20}
\frac{\prt\hat H}{\prt m}  = -\; \frac{1}{m}\; \hat H \; .
\ee
From Eq. (14), we find
\be
\label{21}
\frac{\prt S}{\prt m} = -\bt^2\; {\rm cov}\left (\hat H,\;
\frac{\prt\hat H}{\prt m}\right ) = \frac{\bt^2}{m}\;\Dlt^2(\hat H)\; .
\ee
Remembering the definition of the specific heat
\be
\label{22}
C_V \equiv \frac{1}{N} \left ( \frac{\prt E}{\prt T}\right )_V =
\frac{\Dlt^2(\hat H)}{NT^2} \; ,
\ee
we get
\be
\label{23}
\frac{\prt S}{\prt m} = \frac{1}{m}\; NC_V \; .
\ee
The latter equation emphasizes again that the adiabatic approximation, 
in the sense of the independence of the entropy from the mass, can become
asymptotically exact solely for $T\ra 0$, when $C_V\ra 0$. But for finite 
temperatures, when the specific heat can be rather large, one cannot 
neglect the dependence of $S$ on $m$, so that $\prt S/\prt m\neq 0$.

Finally, from Eq. (17) we find
\be
\label{24}
\frac{\prt E}{\prt m} = \frac{1}{m}\left ( NT C_V - E\right ) \; .
\ee
At zero temperature, we have
\be
\label{25}
\frac{\prt E}{\prt m} = - \; \frac{E}{m} \qquad (T=0) \; ,
\ee
where the limit
$$
\lim_{T\ra 0}\; TC_V = 0
$$
is taken into account.

It is formula (24), instead of Eq. (5), which is to be used, when no 
adiabatic approximation is valid and the potential-energy operator depends 
on the mass according to Eq. (19). At zero temperature, Eq. (24) reduces to 
Eq. (25). For the low-temperature dilute Bose gas, for which $E/N=2\pi\rho 
a_s/m$, Eq. (25) becomes a simple identity. In this way, taking the mass 
derivative correctly, without forgetting the dependence on the mass of the 
potential-energy term, results just in an identity, but by no means this
defines the kinetic energy.

The situation diversifies when, in addition to the interaction energy, 
the potential energy term contains an external potential depending on 
the particle mass. For instance, the external potential can be a 
confining potential acting on atoms in a trap. Let the potential-energy 
operator consist of two terms,
\be
\label{26}
\hat W = \hat V +\hat U \; ,
\ee
of which the first term corresponds to particle interactions, with the 
mass dependence according to Eq. (19), while the second term describes a 
power-law trapping potential, proportional to $m^{n/2}$. For a parabolic
trap, $n=2$. Thus, we have
\be
\label{27}
\frac{\prt\hat V}{\prt m} = -\; \frac{1}{m}\; \hat V \; , \qquad
\frac{\prt\hat U}{\prt m} = \frac{n}{2m}\; \hat U \; .
\ee
Consequently, instead of Eq. (19), we get
\be
\label{28}
\frac{\prt\hat W}{\prt m} = -\; \frac{1}{m}\; \hat W + \frac{1}{m}
\left ( 1 + \frac{n}{2}\right ) \hat U \; .
\ee
Then relation (17) transforms to
\be
\label{29}
\frac{\prt E}{\prt m} = -\; \frac{1}{m}\; E + \frac{1}{m}
\left ( 1 + \frac{n}{2}\right )<\hat U> + \;
T\; \frac{\prt S}{\prt m} \; .
\ee
For a harmonic trap at zero temperature, we obtain
\be
\label{30}
\frac{\prt E}{\prt m} = -\; \frac{1}{m}\; E + \frac{2}{m}\;
<\hat U> \qquad (T=0) \; .
\ee
This again does not allow one to get in a simple way the average kinetic 
energy.

Thus, in any case, considering whether homogeneous systems or trapped 
atoms, one must accurately take into account the mass dependence of the 
potential-energy term. Then no paradoxes appear. While blindly applying 
Eq. (5) outside of the region of its validity yields principally wrong 
results.

It is worth emphasizing that Eq. (25) is an identity, for the dilute Bose 
gas at zero temperature, for all known orders of the perturbation theory 
in powers of the dimensionless coupling $\al\equiv\rho a_s^3\ll 1$. The 
ground-state energy, as an expansion with respect to this coupling, is
$$
\frac{E}{N} = 2\pi\; \frac{\rho a_s}{m} \left ( 1 + b_1\al^{1/2} + 
b_2\al +b_2'\al\ln\al\right ) \; .
$$
The coefficients in this expansion, found in Refs. [11--14] and 
summarized in the review article [15], are
$$
b_1\cong 4.814 \; , \qquad b_2 \cong 71.337\; , \qquad 
b_2'\cong 19.654 \; .
$$
As is evident, Eq. (25) is an identity for the ground-state energy of 
the dilute Bose gas in all orders with respect to $\al$. Moreover, the 
above expansion can be extrapolated, by invoking resummation techniques, 
to arbitrary values of $\al\in[0,\infty)$. Consequently, relation (25) 
seems to be valid, not merely for the dilute Bose gas in the weak-coupling 
limit, but for this gas with arbitrarily strong couplings. 

\section{Kinetic energy of Bose-condensed systems}

The direct definition of the kinetic energy per particle, for a homogeneous 
Bose-Einstein condensed system, is
\be
\label{31}
\overline K \equiv \frac{<\hat K>}{N} = \frac{1}{N} \; \sum_{k\neq 0}\;
\frac{k^2}{2m}\; n_k \; ,
\ee
where $n_k$ is the momentum distribution of atoms. The contribution of
condensed particles is, as is evident, zero because of the zero momenta
of particles in the condensate. For a dilute Bose gas, the number of 
noncondensed particles, with a momentum $k$, is much smaller than 
the number of condensed particles, $n_k\ll N_0\approx N$. The potential
energy is mainly due to the interaction of the condensed atoms, so that 
$<\hat W>\cong 2\pi\rho a_s/m$. But the kinetic energy is caused by the 
noncondensed particles, because of which it must be of the higher order 
with respect to the interaction intensity $\Phi_0=4\pi a_s/m$, hence of 
the higher order in the scattering length $a_s$. From here, even with no 
calculations, it is obvious that the average kinetic energy $<\hat K>$ 
cannot be of the same order in $a_s$ as the ground-state energy 
$E\approx<\hat W>$. Recall that incorrectly employing Eq. (5) one would 
get $<\hat K>\approx E\approx<\hat W>$, which is in an explicit 
contradiction with the direct definition (31).

Keeping in mind a low-temperature dilute Bose gas, satisfying the 
Bogolubov theory [2--5], we have
\be
\label{32}
n_k = \frac{\om_k}{2\ep_k}\; {\rm coth}\left ( \frac{\ep_k}{2T}\right ) -
\; \frac{1}{2} \; .
\ee
Here the notation
$$
\om_k \equiv \frac{k^2}{2m} + mc^2
$$
is used; the Bogolubov spectrum is
$$
\ep_k =\sqrt{(ck)^2 +\left ( \frac{k^2}{2m}\right )^2} \; ,
$$
and the sound velocity is
$$
c\equiv \sqrt{\frac{\rho}{m}\; \Phi_0} = \frac{\sqrt{4\pi\rho a_s}}{m}\; .
$$

The kinetic energy per particle (31) can be represented as
\be
\label{33}
\overline K = \frac{1}{\rho} \; \int \; \frac{k^2}{2m} \; n_k\; 
\frac{d\bk}{(2\pi)^3} \; .
\ee
For the local interactions $\Phi(\br)=\Phi_0\dlt(\br)$, the integrand in 
$\overline K$ behaves as $k^4 n_k\propto\Phi_0^2$ at large $k\ra\infty$, 
hence, the above integral linearly diverges. However, this by no means 
implies the failure of the Bogolubov theory, but simply reflects a defect 
in the dilute-gas approximation. The Bogolubov theory [2--5] has been 
formulated for arbitrary interaction potentials $\Phi(\br)$, but not solely 
for the local potentials. For a realistic interaction potential $\Phi(\br)$, 
its Fourier transform $\Phi_k$ steeply diminishes with increasing $k\ra
\infty$. Therefore, the integrand in $\overline K$ behaves as $k^4n_k\propto
\Phi_k^2$, which is a rapidly diminishing function of $k\ra\infty$. Thus, for 
realistic interaction potentials, no infrared divergence arises in calculating 
$\overline K$. Instead of resorting to realistic interaction potentials, which
would essentially complicate calculations, one can model the fast decay of 
$\Phi_k$ at large $k$ by the step function
\be
\label{34}
\Phi_k = \Phi_0\Theta(k_0-k) \; ,
\ee
with $\Theta(k)$ being the unit step function. The boundary wave vector 
$k_0$ can be chosen as the vector, where the phonon behaviour of the 
Bogolubov spectrum $\ep_k$ changes to the single-particle type dependence. 
This definition gives $k_0=2mc$.

Our goal here is to evaluate the mean kinetic energy (33) by its order of 
magnitude, in order to demonstrate its nonlinear dependence on $a_s$, which 
makes it principally different from the ground-state energy, whose main term 
is linear in $a_s$. To this end, we resort to the simple form (34), assuming 
that, by the order of magnitude, the integral (33) can be evaluated by 
cutting off its upper limit with the boundary wave vector. Let us also 
consider the zero-temperature case, when the particle distribution (32) 
becomes
$$
n_k = \frac{\om_k-\ep_k}{2\ep_k} \qquad (T=0) \; .
$$
Then the integral (33) gives
\be
\label{35}
\overline K = C\; \frac{\rho a_s}{m}\; \sqrt{\rho a_s^3} \; ,
\ee
with the constant
$$
C= \frac{64}{5}\left ( 3\sqrt{2} \; - 4\right )\sqrt{\pi} \cong 5.505 \; .
$$
As is expected, $\overline K\propto a_s^{5/2}$, making it principally 
different from $E\propto a_s$.

In this way, the Landau-Lifshitz equation (5) is an adiabatic approximation,
valid under conditions (1)-(3), together with the assumption $\prt S/\prt 
m=0$. Incorrectly using this relation, outside its region of applicability, 
leads to wrong paradoxical results. For instance, condition (3) does not 
hold true for the dilute Bose gas, whose interaction potentials do depend 
on particle mass.

Generally, a parametric differentiation for the statistical average of an 
operator is given by formula (7). If the operator is a self-adjoint integral 
of motion, it satisfies the zero-temperature limit (9).

The parametric differentiation of the internal energy is represented by 
the general expressions (10), (12), and (13). The mass differentiation of 
the internal energy is given by formula (17). For the dilute Bose gas, the 
latter takes the form (24), which at zero temperature reduces to relation 
(25). In the case of trapped atoms, the mass variation of the internal 
energy leads to Eq. (29). For the homogeneous Bose gas, the mean kinetic 
energy per particle can be evaluated as in Eq. (35). The dependence on the 
scattering length of the kinetic energy $\overline K\propto a_s^{5/2}$ and
of the internal energy $E\propto a_s$ is fundamentally different.

\end{document}